# Normative Epistemology for Lethal Autonomous Weapons Systems

S. Kate Devitt[1,2]

[1]Trusted Autonomous Systems

[2]University of Queensland

Corresponding author: kate.devitt@tasdcrc.com.au

## Abstract

The rise of human-information systems, cybernetic systems, and increasingly autonomous systems requires the application of epistemic frameworks to machines and human-machine teams. This chapter discusses higher-order design principles to guide the design, evaluation, deployment, and iteration of Lethal Autonomous Weapons Systems (LAWS) based on epistemic models. Epistemology is the study of knowledge. Epistemic models consider the role of accuracy, likelihoods, beliefs, competencies, capabilities, context, and luck in the justification of actions and the attribution of knowledge. The aim is not to provide ethical justification for or against LAWS, but to illustrate how epistemological frameworks can be used in conjunction with moral apparatus to guide the design and deployment of future systems. The models discussed in this chapter aim to make Article 36 reviews of LAWS systematic, expedient, and evaluable. A Bayesian virtue epistemology is proposed to enable justified actions under uncertainty that meet the requirements of the Laws of Armed Conflict and International Humanitarian Law. Epistemic concepts can provide some of the apparatus to meet explainability and transparency requirements in the development, evaluation, deployment, and review of ethical AI.

## Introduction

In Army of None, Paul Scharre (2018) tells the story of a little Afghan goat-herding girl who circled his sniper team, informing the Taliban of their location via radio. Scharre uses this story as an example of a combatant who he, and his peers did not target—but that a lethal autonomous weapon programmed to kill, might legally target. A central mismatch between humans and robots, it seems, is that humans know when an action is right, and a robot does not. In order for any Lethal Autonomous Weapon Systems (LAWS) to be ethical, it would need—at a minimum—to have situational understanding, to know right from wrong, and to be operated in accordance with this knowledge.

Lethal weapons of increasing autonomy are already utilized by militant groups, and they are likely to be increasingly used in Western democracies and in nations across the world (Arkin et al. 2019; Scharre 2018). Because they will be developed—with varying degrees of human-in-the-loop—we must ask what kind of design principles should be used to build them and what should direct that design? To trust LAWS we must trust that they know enough about the world, themselves, and their context of use to justify their actions. This chapter interrogates the concept of knowledge in the context of LAWS. The aim of the chapter is not to provide an ethical framework for their deployment, but to illustrate epistemological frameworks that could be used in conjunction with moral apparatus to guide the design and deployment of future systems.

"Epistemology" is the study of knowledge (Moser 2005; Plato 380 B.C.; Quine 1969). Traditionally conceived, epistemology is the study of how humans come to know about the world via intuition, perception, introspection, memory, reason, and testimony. However, the rise of human-information systems, cybernetic systems, and increasingly autonomous systems requires the application of epistemic frameworks to machines and human-machine teams. Epistemology for machines asks the following: How do machines use sensors to know about the world? How do machines use feedback systems to know about their own systems including possible working regimes, machine conditions, failure modes, degradation patterns, history of operations? And how do machines communicate system states to users (Bagheri et al. 2015) and other machines? Epistemology for human-machine teams asks this: How do human-machine teams use sensors, perception, and reason to know about the world? How do human-machine teams share information and knowledge bidirectionally between human and machine? And how do human-machine teams communicate information states to other humans, machines, and systems?

Epistemic parameters provide a systematic way to evaluate whether a human, machine, or human-machine teams are trustworthy (Devitt 2018). Epistemic concepts underpin assessments that weapons do not result in superfluous injury or unnecessary suffering, weapons systems are able to discriminate between combatants and non-combatants, and weapons effects are controlled (Boothby 2016). The models discussed in this chapter aim to make Article 36 reviews of LAWS (Farrant and Ford 2017) systematic, expedient, and evaluable. Additionally, epistemic concepts can provide some of the apparatus to meet explainability and transparency requirements in the development, evaluation, deployment, and review of ethical AI (IEEE Global Initiative on Ethics of Autonomous and Intelligent Systems 2019; Jobin et al. 2019).

Epistemic principles apply to technology design, including automatic, autonomous, and adaptive systems, including how an artificial agent ought to modify and develop their own epistemic confidences as they learn and act in the world. These principles also guide system designers on how autonomous systems must communicate their doxastic states to humans. A doxastic state relates its agent's attitude to its beliefs, including its confidence, uncertainty, skepticism, and ignorance (Chignell 2018). Systems must communicate their representations of the world and their attitudes to these representations. Designers must program systems to act appropriately under different doxastic states to ensure trust. Before setting off, I would like to acknowledge the limits of my own knowledge with regard to epistemology. This chapter will draw on the Western philosophical tradition in which I am trained, but I want to be clear that this is only one of many varied epistemologies in a wide range of cultural traditions (Mizumoto et al. 2018). The reader (and indeed the author) is recommended to explore alternate epistemologies on this topic such as ethnoepistemology (Maffie 2019) and the Geography of Philosophy project (Geography of Philosophy Project 2020). Additionally, there is a wide corpus of literature on military leadership that is relevant to epistemic discussions, particularly developing virtuous habits with regards to beliefs (e.g., Paparone and Reed 2008; Taylor et al. 2018). Note, I will not be addressing many ethical criticisms of LAWS such as the dignity argument (that death by autonomous weapons is undignified) or the distance argument (that human involvement in drone warfare is too psychologically and spatially distant for operators to act morally when LAWS are under their control).

This chapter, by introducing some foundational concepts and exploring a subset of epistemologies that will hopefully add to the debate on lethal autonomous weapons by highlighting the need for higher-order epistemic models to guide not only the design but also the training, evaluation, and regulation of lethal autonomous weapons systems irrespective of specific techniques and technologies used in their creation.

## Motivation

Meaningful human control is a critical concept in the current debates on how to constrain the design and implementation of LAWS. The 2018 Group of Governmental Experts on Lethal Autonomous Weapons Systems (LAWS) argued that,

In the context of the deployment and use of a weapons system in an armed conflict, delegations noted that military personnel activate the weapons systems and monitor their functioning. This would require that the operator know the characteristics of the weapons system, is assured that they are appropriate to the environment in which it would be deployed and has sufficient and reliable information on them in order to make conscious decisions and ensure legal compliance. . . . [P]ositive measures are necessary to prevent indiscriminate action and injury by lethal autonomous weapons systems caused by a breakaway from human control. To develop such measures, concepts such as 'meaningful human control' and 'human judgment' need to be further elaborated and clarified. (Annex III, 18 CCW GGE LAWS 2018: 14)

"Meaningful human control" requires operators to know that systems are reliable and contextually appropriate. How should the parameters of human control and intervention be established? Researchers are grappling with the theoretical and practical parameters of meaningful human control to inform design criteria and legal obligation (Horowitz and Scharre 2015; Santoni de Sio and van den Hoven 2018). Santoni de Sio and van den Hoven (2018) argue that the analysis should be informed by the concept of "guidance control" from the free will and moral responsibility literature (Fischer and Ravizza 1998) translated into requirements for systems, engineering, and software design (van den Hoven 2013). In order to be morally responsible for an action X, a person or agent should possess "guidance control" over that action. Guidance control means that the person or agent is able to reason through an action in the lead-up to a decision, has sufficient breadth of reasoning capability to consider the ethical considerations (and change their actions on the basis of ethical considerations), and uses its own decisional mechanisms to make the decision.

A morally responsible agent has sensitive and flexible reasoning capabilities, is able to understand that its own actions affect the world, is sensitive to others' moral reactions toward it, and is in control of its own reasoning as opposed to being coerced, indoctrinated, or manipulated. An argument to ban LAWS is that a decision by a LAWS to initiate an attack—without human supervision—in an unstructured environment is not under meaningful human control because LAWS cannot track the reasoning required by international law (including necessity, discrimination, and proportionality), and they are not adaptive to changing morally relevant features of their operational environment (Asaro 2012; Sharkey 2012).

It is worth considering meaningful human control in terms of the trajectory of an action, from decision into consequences in the world. An expert archer shooting an arrow at a target has meaningful human control when they use their competence to set up the shot, to sense the changing wind, and fire during a lull. A novice archer is not in meaningful control because they lack the capacity to guide the arrow to the target—even though they may wish to hit the bullseye—and they fire rashly—without appropriate reflection on their own state or the environmental conditions that might affect their accuracy. The novice lacks full control. The trajectory of the arrow between the archer firing and it landing on the target is constrained both by the energy imbued within it from the archer and the vicissitudes of the wind that may blow it off course. Meaningful control over a projectile is affected by the amount of time it takes from a human making a decision to the outcomes and consequences of that decision and the uncertainties that could affect the outcome during the intervening period.

If an arrow was equipped with autonomous flight stabilizers to reorient itself to its human-programmed trajectory, then it is not controlled by a human, but it is operating in abidance with human intent. This is how Tomahawk cruise missiles and loitering munitions are programmed to persist to a human-selected target (Gettinger and Michel 2017; Raytheon 2019a). If the moral landscape changes in the time between launch and effect, for most of the history of weapons that operate at a distance, the shot lands and collateral damage is accepted as a terrible but not immoral consequence of the human action—because humans operated to the best of their capability and knowledge. Put another way, the projectile missed the legitimate target not due to human incompetence, but due to the manifestation of an unforeseen uncertainty, an unlucky intervention.

Suppose the collateral damage was a child walking into a targeted building. Would it be moral if the projectile was able to autonomously abort their mission, adjust its payload, or alter its trajectory on the basis of new information about the decision?—preventing harms seems just. What if the collateral damage of hitting one corner of the building could be calculated and contrasted with the collateral damage of hitting the other corner of the building faster than human thought for a time-critical mission. Could an autonomous weapon reason through the choice of which corner to strike to reduce harms? If the reasoning was robust, the ends seem to justify it. But, would this system still be in meaningful human control according to International Humanitarian Law (IHL)? Humans made the decision about the time sensitivity of the target in order to authorize the LAWS to adapt its target. Systems may be able to reason through many aspects of distinction and proportionality and eventually perhaps even necessity faster and more comprehensively than a human. Their timely knowledge advantage is a reason to consider their use as ethically justified.

Perhaps there are decisions for which humans should not be in meaningful control? As Hancock (2016: 289) asks,

> can we truly say that we always know best? If we examine our current record of planetary stewardship it is painfully obvious that we are lacking in both rationality and a necessary standard of care. It may well be possible that globally-interconnected operations are better conducted by quasi- and subsequently fully autonomous systems?

How many unjust harms could be mitigated with systems more knowledgeable than humans engaged in guarding civilians and defending protected objects (Scholz and Galliott 2018)? What is relevant for this chapter is not to settle the debate on meaningful human control or abidance with IHL, but to illustrate the role of knowledge in the evaluation and acceptability of LAWS. In particular, it is about motivating the study of epistemic frameworks to understand why a particular deployment of LAWS may be deemed unethical due to it lacking the relevant knowledge to justify autonomous decision-making.

The structure of the chapter is as follows. First I will introduce the nuts and bolts of epistemology including the analysis of knowledge as 'justified true belief' and what that entails for different doxastic states of agents in conflict that might justify actions. I introduce cases where conditions of knowledge are threatened, called 'Gettier Cases' and explain why they are relevant in conflicts where parties are motivated to deceive. I will then work through the cognitive architecture of humans and artificial agents within which knowledge plays a functional role. The degree to which humans might trust LAWS in part depends on understanding how LAWS make decisions including how machines form, update and represent beliefs and how beliefs influence agent deliberation and action selection. The chapter finishes with a discussion of three normative epistemological frameworks: reliabilism, virtue epistemology and Bayesian epistemology. Each of these offer design frameworks that LAWS can be evaluated against in their performance of tasks in accordance with commander's intent, IHL and Laws of Armed Conflict (LOAC).

# Epistemology

Epistemology is the study of how humans or agents come to know about the world. Knowledge might be innate (hardwired into a system), introspected, or learned (like Google maps dynamically updating bushfire information). The dominant conception of knowledge in the Western tradition is that a human knows p when she has justified true belief that p (Plato 369 B.C.). Under this framework, a Combatant (Ct) knows that the Person (P) is a Civilian (Cv) not a Belligerent (B) if the following are present.

## Justified True Belief

   a) Ct accurately identifies P as Cv,

   b) Ct believes that they have accurately identified P as Cv, and

   c) Ct is justified in believing that they have accurately identified P as Cv.

The enterprise of epistemology has typically involved trying to understand (1) What is it about people that enable them to form beliefs that are accurate, for example, what enables a combatant to identify a civilian versus a lawful target? (2) When are people justified in holding certain beliefs, for example, under what conditions are identification attributes in play and defensible? and (3) What warrants this justification, for example, what features about human perception, mission briefings, ISR, patterns of behaviour, command structures, and so forth enable the combatant to have justified true beliefs? To better grasp the explanatory usefulness of knowledge as justified true belief, we can explore conditions where a combatant has false beliefs, does not believe, or has unjustified beliefs in order to tease out these component parts. Let's consider the same situation under the false belief, disbelief, and unjustified belief scenarios.

## False Belief

   a) Ct inaccurately identifies P as Cv instead of B,

   b) Ct believes that they have accurately identified P as Cv, and

   c) Ct is not justified in believing that they have accurately identified P as Cv.

False belief is the most significant threat to the combatant, as the misidentified belligerent may take advantage of the opportunity to aggress. If the combatant comes to harm, a post-event inquiry might find the combatant's own perceptual judgment faulty, or that the belligerent was able to conceal their status, such as by not wearing a uniform, traveling with a child, hiding their weapons, or spoofing intelligence networks.

## Disbelief

a) Ct accurately identifies P as Cv,

b) Ct does not believe that they have accurately identified P as Cv, and

c) Ct is justified in believing that they have accurately identified P as Cv.

Disbelief may occur in "the fog of war," where evidence sufficient to justify a combatant's belief that a person is a civilian does not in fact influence their beliefs. Perhaps intelligence has identified the person as a civilian via methods or communication channels not trusted by the combatant—they may worry they're being spoofed. Or they doubt their own perceptual facilities—perhaps there really is a rifle inside the person's jacket? Disbelief can have multiple consequences; a cautious but diligent combatant may seek more evidence to support the case that the person is a civilian. Being cautious may have different impacts depending on the tempo of the conflict and the time criticality of the mission. If the combatant develops a false belief that the person is a belligerent, it may lead them to disobey IHL.

## Unjustified Belief

a) Ct accurately identifies P as Cv,

b) Ct believes that they have accurately identified P as Cv, and

c) Ct is not justified in believing that they have accurately identified P as Cv.

Unjustified belief occurs when the combatant believes that the person is a civilian by accident, luck, or insufficient justification rather than via systematic application of reason and evidence. For example, the combatant sees the civilian being assisted by Médecins Sans Frontières. The person really is a civilian, so the belief is accurate, but it arose through unreliable means because both combatants and civilians are assisted by Médecins Sans Frontières (2019). The protection of civilians and the targeting process should not be accidental, lucky, or superficial. Combatants should operate within a reliable system of control that ensures that civilians and combatants can be identified with comprehensive justification—and indeed, many Nations abide by extensive systems of control, for example, the Australian government (2019). Consider a LAWS Unmanned Aerial Vehicle (UAV) that usually has multiple redundant mechanisms for identifying a target—say autonomous visual identification and classification, human operator, and ISR confirmation. If communication channels were knocked out so that UAV decisions were based on visual feed and internal mechanisms only, it might not be sufficient justification for knowledge, and the UAV may have to withdraw from the mission.

## Gettier Cases

Gettier cases occur in cases when achieving justified true belief is not said to produce knowledge (Gettier 1963). Imagine the combatant in a tank, receiving data on their surrounding environment through an ISR data feed that is usually reliable. They are taking aim at a building, a military target justified by their ROE and IHL. The data feed indicates that there are people walking in front of the building. Suppose the ISR feed was intercepted by an adversary, and this intercept went undetected by the combatant or their command. The false data feed is designed to trick the combatant, so warns them that civilians would be harmed if she takes the shot. As it happens, there are civilians walking past the building during the same time as the combatant is receiving the manipulated data feed. In this case, the following are possible.

## Gettier Case

a) Ct accurately identifies P as Cv (because civilians are walking past the building),

b) Ct believes that they have accurately identified P as Cv, and

c) Ct is justified in believing that they have accurately identified P as Cv (because the ISR feed has been accurate and reliable in the past).

Epistemologists have created a "Gettier industry" to try and develop conceptions of knowledge that both explain and triumph over Gettier cases (Lycan 2006; O'Brien 2016). I am unable to cover all responses, but I will explore several epistemic models useful for designing LAWS and worthy of further investigation. Two approaches for solving Gettier cases are (1) the addition of different conditions for knowledge and (2) not requiring knowledge to obtain to justify belief. I explore both approaches in the following sections.

## Tracking

Nozick (1981) claims that Gettier cases can be resolved if the right sort of causal structure is in play for the attribution of knowledge. He calls this concept "tracking." Nozick supposes that to know, we must be using our facilities to track the structure of the world so that we are sensitive to factors that might affect knowledge attainment. So, for Nozick, for knowledge to occur for the combatant in the tank, the following conditions must be met:

a) P is a Cv (The person is a civilian);

b) Ct believes that P is a Cv (the combatant believes that P is a Cv);

c) If it were not the case that P is a Cv, then Ct would not believe that P is a Cv (if the person was not a civilian, the combatant would not believe that the person was a civilian);

d) If it were the case that P is a Cv, then Ct would believe that P is a Cv (if the person was a civilian, then Ct would believe that the person is a civilian).

In this case, the combatant does not have knowledge because they would still believe that P was a civilian even if P was a combatant or even if P wasn't there at all. The lack of situational awareness of causal factors that affected beliefs explains the lack of knowledge. Adversaries will try to manipulate LAWS to believe the wrong state of the world, and thus significant efforts must be made to ensure LAWS' beliefs track to reality.

## Belief

Core to the construction of knowledge is the concept of belief. It is worth being clear on what a belief is and what it does in order to understand how it might operate inside an autonomous system. In this chapter, I take "belief" to be a propositional attitude like "hope" and "desire," attached to a proposition such as "there is a hospital." So, I can believe that there is a hospital, but I can also hope that there is a hospital because I need medical attention. Beliefs can be understood in the functional architecture of the mind as enabling a human or an agent to act, because it is the propositional attitude we adopt when we take something to be true (Schwitzgebel 2011). In this chapter, beliefs are treated as both functional and representational. Functionalism about mental states is the view that beliefs are causally related to sensory stimulations, behaviours, and other mental states (Armstrong 1973; Fodor 1968). Representationalism is the view that beliefs represent how things stand in the world (Fodor 1975; Millikan 1984).

Typically, the study of knowledge has assumed that beliefs are all or nothing, rather than probabilistic (BonJour 1985; Conee and Feldman 2004; Goldman 1986). However, Pascal and Fermat argued that one should strive for rational decision-making, rather than truth—the probabilistic view (Hajek and Hartmann 2009). The all-or-nothing and probabilistic views are illustrated by the difference between Descartes's Cogito and Pascal's wager. In the Cogito, Descartes argues for a belief in God based on rational reflection and deductive reasoning. In the wager, Pascal argues for a belief in God based on outcomes evaluated probabilistically. Robots and autonomous systems can be built with a functional architecture that imitates human belief structures called belief/desire/intention models (Georgeff et al. 1999). But, there are many cognitive architectures that can be built into a robot, and each may have different epistemic consequences and yield different trust relations (Wagner and Briscoe 2017). Mental states and mental models can be developed logically to enable artificial systems to instantiate beliefs as propositions (Felli et al. 2014) or probabilities (Goodrich and Yi 2013). The upshot is that the functional role of belief for humans can be mimicked by artificial systems—at least in theory. However, it is unclear whether the future of AI will aim to replicate human functional architecture or approach the challenge of knowledge quite differently.

## The Representation of Knowledge

How is knowledge represented and operated on within humans, autonomous systems, and human-machine teams? The ancient Greeks imagined that perceptual knowledge was recorded and stored in the mind, similarly to the way wax takes on the imprint of a seal (Plato 369 B.C.). Early twentieth-century thinkers used the metaphor of modern telephony, radio, and office workers to explain how signals from the world translated into thoughts, and thoughts were communicated between different brain regions (Draaisma 2000). The late twentieth century saw the rise of the computational model of the mind, supposing that humans store knowledge as representations akin to modern symbolic computers—allowing functions such as combining propositions using a language of thought and retrieving memories from storage (Fodor 1975). In the 1980s, the connectionist view of the mind arose based on neuroscience. Under connectionism, the mind is identified with the biological and functional substrate of the brain and simulated by building artificial neural networks (ANN) (Churchland 1989; Fodor and Pylyshyn 1988; Kiefer 2019). Connectionist networks such as deep learning succeed at tasks via networks of neurons arranged in a hierarchy. Each object, concept, or thought can be represented by a vector (LeCun 2015):

i.    [-0.2, 0.3, -4.2, 5.1, . . .] represent the concept "cat";
ii.   [-0.2, 0.4, -4.0, 5.1, . . .] represent the concept "dog."

Vectors, i) and ii) are similar because cats and dogs have many properties in common. Reasoning and planning consist of combining and transforming thought vectors. Vectors can be compared to answer questions, retrieve information, and filter content. Thoughts can be remembered by adding memory networks (simulating the hippocampal function in humans) (Weston et al. 2014). Grouping neurons into capsules allows retention of features of lower-level representations as vectors are moved up the ANN hierarchy (Lukic et al. 2019). Contrary to early critiques of "dumb connectionist networks," the complex structures and functionality of contemporary ANN meet the threshold required of sophisticated cognitive representations (Kiefer 2019)—although not yet contributing to explanations of consciousness, affect, volition, and perhaps reflective knowledge. Humans and technologies can represent the world, make judgments about propositions, and can be said to believe when they act as though propositions were true.

Knowledge might be about facts, but it is also about capabilities, the distinction between "knowing that" versus "knowing how" (Ryle 1949). Knowing that means the agent has an accurate representation of facts, such as the fact that medical trucks have red crosses on them, or that hospitals are protected places. Knowing how are skills, such as the ability to ride a bike, the ability of a Close in Weapons System (CIWS) to reliably track incoming missiles or loitering munition's ability to track a human-selected target (Israel Aerospace Industries 2019; Raytheon 2019b). Knowledge how describes the processes used to able humans or agents to perform tasks such as manoeuvring around objects or juggling that may or may not be propositional in nature. Knowledge how can be acquired via explicit instruction, practice, and experience and can become an implicit capability over time. The reduction of cognitive load when an agent moves from learner to expert explains the gated trajectory through training programs for complex physical tasks such as military training. Examples of knowledge how include the autopilot software on aircraft, or AI trained how to navigate a path or play a game regardless of whether specific facts are represented at any layer. An adaptive autonomous system can learn and improve knowledge how in situ. Any complex systems incorporating LAWS, such as human-autonomy teams (Demir et al. 2018) and manned-unmanned teams (Lim et al. 2018) must be assessed for how the system knows what it is doing and how it knows ethical action. Knowledge how is what a legal reviewer needs to assess in order to be sure LAWS are compliant with LOAC (Boothby 2016). Knowledge how is pertinent to any ethical evaluative layer, or ethical "governor" (Arkin et al. 2009). Evaluating knowledge how requires a testing environment simulating multiple actions, evaluating them against ethical requirements (Vanderelst and Winfield 2018). Many of the capabilities of LAWS are perhaps best understood as knowledge how rather than knowledge that, and our systems of assurance must be receptive to the right sort of behavioural evidence for their trustworthiness.

A concern for Article 36 reviews of LAWS is that they are a black box—that the precise mechanisms that justify action are hidden or obtuse to human scrutiny (Castelvecchi 2016).

Not even AlphaGo's developer team is able to point out how AlphaGo evaluates the game position and picks its next move (Metz 2016). Three solutions emerge from the black box criticism of AI: (1) Unexplainable AI is unethical and must be banned; (2) Unexplainable AI is unethical, and yet we need to have it anyway; and (3) Unexplainable AI can be ethical under the right framework.

To sum up, so far I have examined the conditions of knowledge (justified true belief), belief, how beliefs are represented and used to make decisions. I now move to normative epistemology, theories that help designers of LAWS to ensure AI and artificial agents are developed with sufficient competency to justify their actions in conflict. Given the 'blackbox' issues with some artificial intelligence programming, I argue that reliabilism is an epistemic model that allows for systems to be tested, evaluated, and trusted despite some ignorance with regards to how any specific decision is made.

## Reliabilism

Skeptical arguments show that there are no necessary deductive or inductive relationships between beliefs and their evidential grounds or even their probability (Greco 2000). In order to avoid skepticism, a different view of what constitutes good evidence must be found. Good evidence for a positive epistemology might be the reliable connection between what we believe about the world, and the way the world behaves (which is consistent with these beliefs), such that "the grounds for our beliefs are reliable indications of their truth" (Greco 2000, 4). Reliabilism supposes that a subject knows a proposition p when (a) p is true, (b) the subject believes p, and (c) the belief that p is the result of a reliable process. A key benefit of reliabilism is that beliefs formed reliably have epistemic value, regardless of whether an agent can justify or infer reasons for their reliability. Reliable beliefs, like the readings from a thermometer or thermostat, are externally verifiable. Cognitive agents are more complex than thermometers, however. Agents have higher-order reliability based on the reliability of subsystems. In complex agents, the degree of reliability of the process gives the degree to which a belief generated by it is justified (Sosa 1993). The most discussed variant of reliabilism is process (or paradigm) reliabilism: "S's belief in p is justified if it is caused (or causally sustained) by a reliable cognitive process, or a history of reliable processes" (Goldman 1994). An issue for reliabilists is a lack of sophistication about how cognitive processes actually operate—a similar problem for the internal processes of AI build by machine learning and deep learning.

15.9:    Virtue Epistemology

Virtue epistemology is a variant of reliabilism in which the cognitive circumstances and abilities of an agent play a justificatory role.  In sympathy with rationalists (Descartes 1628/1988, 1–4, Rules II and III; Plato 380 B.C.), virtue epistemologists argue that there is a significant epistemic project to identify intellectual virtues that confer justification upon a true belief to make it knowledge. However, virtue epistemologists are open to empirical pressure on these theories. Virtue epistemology aims to identify the attributes of agents that justify knowledge claims. Like other traditional epistemologies, virtue epistemology cites normative standards that must be met in order for an agent's doxastic state to count as knowledge, the most important of which is truth. Other standards include reliability, motivation, or credibility. Of the many varieties of virtue epistemology (Greco 2010; Zagzebski 2012), I focus on Ernie Sosa's that specifies an iterative and hierarchical account of reliabilist justification (Sosa 2007; 2009; 2011), particularly useful when considering nonhuman artificial agents and the doxastic state of human-machine teams. Sosa's virtue epistemology considers two forms of reliabilist knowledge: animal and reflective.

## Animal Knowledge

Animal knowledge is based on an agent's capacity to survive and thrive in the environment regardless of higher-order beliefs about its survival, without any reflection or understanding (Sosa 2007). An agent has animal knowledge if their beliefs are accurate, they have the skill (i.e., are adroit) at producing accurate beliefs, and their beliefs are apt (i.e., accurate due to adroit processes). Consider an archer shooting an arrow at a target. A shot is apt when it is accurate not because of luck or a fortuitous wind that pushes the arrow to the center, but because of the competence exhibited by the archer. Similarly an  autonomous fire-fighting drone is apt when fire retardant is dropped on the fire due to sophisticated programming, and comprehensive test and evaluation.  Sosa takes beliefs to be long-sustained performances exhibiting a combination of accuracy, adroitness, and aptness. Apt beliefs are accurate (true), adroit (produced by skillful processes), and are accurate because they are adroit. Aptness is a measure of performance success, and accidental beliefs are therefore not apt, even if the individual who holds those beliefs is adroit. Take, for example, a skilled archer who hits the bullseye due to a gust of wind rather than the precision of his shot. Animal knowledge involves no reflection or understanding. However, animal knowledge can become reflective if the appropriate reflective stance targets it. For example, on one hand, a person might have animal knowledge that two combatants are inside an abandoned building, and when questioned, they reflect on their belief and form a reflective judgment that the people are combatants is in an abandoned building with the addition of explicit considerations of prior surveillance of this dwelling, prior experience tracking these combatants, footprints that match the boot tread of the belligerents' uniform, steam emerging from the window, and so forth. On the other hand, animal knowledge might "remain inarticulate" and yet yield "practically appropriate inferences" nevertheless, such as fighter pilot knowledge of how to evade detection, developed through hours of training and experience without the capacity to enunciate the parameters of this knowledge. The capacity to explain our knowledge is the domain of reflective knowledge.

## Reflective Knowledge

Reflective knowledge is animal knowledge plus an "understanding of its place in a wider whole that includes one's belief and knowledge of it and how these come about" (Kornblith 2009, 128). Reflective knowledge draws on internalist ideas about justification (e.g., intuition, intellect, and so on) in order to bolster and improve the epistemic status brought via animal knowledge alone. Reflective knowledge encompasses all higher-order thinking (metacognition), including episodic memory, reflective inference, abstract ideas, and counterfactual reasoning. Animal and reflective knowledge comport with two distinct decision-making systems: (mostly) implicit System 1, and explicit System 2 (Evans and Frankish 2009; Kahneman 2011; Stanovich 1999; Stanovich and West 2000). System 1 operates automatically and quickly, with little or no effort and no sense of voluntary control. System 2 allocates attention to the effortful mental activities that demand it, including complex computations. The operations of System 2 are often associated with the subjective experience of agency, choice, and concentration. System 1 operates in the background of daily life, going hand in hand with animal knowledge. System 1 is activated in fast-tempo operational environments, where decision-making is instinctive and immediate. System 2 operates when decisions are high risk, ethically and legally challenging. System 2 is activated in slow-tempo operational environments where decisions are reviewed and authorized beyond an individual agent.

Virtue epistemology is particularly suited to autonomous systems that learn and adapt through experience. For example, autonomous systems, when first created, may perform fairly poorly and be untrustworthy in a range of contexts—perhaps every context. But, the expectation is that they are trained and deployed in a succession of constrained environments. Constrained environments are built with features and scenarios that AIs learn from, to build their competence and then they are exposed to more complexity and uncertainty as they build decision-making skills. Failure scenarios include unpredictable operating circumstances and operating in adversarial environments designed to trick machine learning algorithms and other experience-based AI. One way to improve decision-making under uncertainty is to allow beliefs to be expressed as a matter of degree, rather than certainty. This is the realm of Bayesian epistemology.

## Bayesian Epistemology

Bayesian epistemology argues that typical beliefs exist (and are performed) in degrees, rather than absolutes, represented as credence functions (Christensen 2010; Dunn 2010; Friedman 2011; Joyce 2010). A credence function assigns a real number between 0 and 1 (inclusive) to every proposition in a set. The ideal degree of confidence an agent has in a proposition is the degree that is appropriate given the evidence and situation the agent is in. No agent is an ideally rational agent, capable of truly representing reality, so they must be programmed to revise and update their internal representations in response to confirming and disconfirming evidence, forging ahead toward ever more faithful reconstructions of reality.

Bayesian epistemology encourages a meek approach with regard to evidence and credences. As Hajek and Hartmann (2009) argue, "to rule out (probabilistically speaking) a priori some genuine logical possibility would be to pretend that one's evidence was stronger than it really was." Credences have value to an agent, even if they are considerably less than 1, and therefore are not spurned. Contrast this with the typical skeptic in traditional epistemology whose hunches, suppositions, and worries can accelerate the demise of a theory of knowledge, regardless of their likelihood. Even better than an abstract theory, the human mind, in many respects, operates in accordance with the tenants of Bayesian epistemology. Top-down predictions are compared against incoming signals, allowing the brain to adapt its model of the world (Clark 2015; Hohwy 2013; Kiefer 2019; Pezzulo et al. 2015).

Bayesian epistemology has several advantages over traditional epistemology in terms of its applicability to actual decision-making. Firstly, Bayesian epistemology incorporates decision theory, which uses subjective probabilities to guide rational action and (like virtue epistemology) takes account of both agent desires and opinions to dictate what they should do. Traditional epistemology, meanwhile, offers no decision theory, only parameters by which to judge final results. Secondly, Bayesian epistemology accommodates fine-grained mental states, rather than binaries of belief or knowledge. Finally, observations of the world rarely deliver certainties, and each experience of the world contributes to a graduated revision of beliefs. While traditional epistemology requires an unforgiving standard for doxastic states, Bayesian epistemology allows beliefs with low credences to play an evidential role in evaluating theories and evidence. In sum, the usefulness of Bayesian epistemology lies in its capacity to accommodate decision theory, fine-grained mental states, and uncertain observations of the world.

A comprehensive epistemology for LAWS will not merely specify the conditions in which beliefs are justified; it will also offer normative guidance for making rational decisions. Virtue epistemology and Bayesian epistemology (incorporating both confirmation theory and decision theory) provide parameters for design of LAWS that explain and justify actions and include a comprehensive theory of decision-making that links beliefs to the best course of action.

## Discussion

Imagine three autonomous systems: AS1, AS2, and AS3.

> AS1: programmed according to virtue epistemology (AS1v),

> AS2: programmed according to Bayesian epistemology (AS2b), and

> AS3: programmed according to Bayesian virtue epistemology (AS3bv).

How will the three be deployed differently?

AS1v (autonomous systems with virtue epistemology) should only act when it knows, which means it must be trained to be highly competent in the specific domain of operation. This makes it highly reliable in a narrow window of operations. Automated countermeasures such as the Phalanx Close In Weapons System (CIWS) are excellent examples of reliable AS1v. When in automated mode, they search for, detect, track, engage, and confirm actions using computer-controlled radar systems. CIWS criteria for targeting include the following: Is the range of the target increasing or decreasing in relation to the ship? Is the contact capable of maneuvering to hit the ship? Is the contact traveling between the minimum and maximum velocities? These actions are safe because only human-made offensive weapons could meet the criteria to make the autonomous weapons system fire. If an incident does occur, then the parameters of the autonomous system are altered to ensure more competent safer operations.

AS2b (autonomous systems with Bayesian epistemology) acts when it has rational belief for action, and this may fall short of knowledge. Consider if AS2b was an autonomous logistics vehicle evaluating their surroundings. It plans an efficient route to a goal, predicts the paths of other objects, creates a map of better and worse paths based on collision avoidance, and updates the route to optimize its path and keep the environment safe (Leben 2019). There are many unknown scenarios for AS2b to manage because there are lots of situations where the robot might collide with objects if they move in a way the system is unable to predict and/or respond to. However, the risk of harmful collision might be deemed very low in a specific and constrained environment; thus it is certified to operate. A Bayesian human combatant operates in highly reliable conditions, but also in the "fog of war." Training, preparation and limited missions reduce the risk of error during lethal actions. But errors do occur, and these are managed so long as combatants have done their best to abide by IHL and the error was an unforeseen occurrence. Could a lethal autonomous weapons system AS2b be deployed similarly to a human combatant? It is likely that LAWS will face an asymmetric evaluation to a human operator. That is to say, the competence of the LAWS will be required to be many times higher if deployed under the same conditions as a human because incorrect task performance will not just affect one unit, but could be instantiated over hundreds or even thousands of implementations of the technology. This point has been made in the autonomous cars literature and is likely to be even more emotive in the regulation of LAWS (Scharre 2018).

AS3bv (autonomous system with Bayesian and virtue epistemology) acts in ways [Bm, Bn …] when it has rational belief for action and [Vm, Vn …] when it has knowledge. Suppose AS3bv is an autonomous drone. AS3bv performs low-risk actions using a Bayesian epistemology such as navigating the skies and AI classification of its visual feed even though it may not exist in a knowledge state. AS3bv considers many sorts of evidence, acknowledges uncertainty, is cautious, and will progress toward its mission even when uncertain. However, when AS3bv switches to a high-risk action, such as targeting with lethal intent, the epistemic mechanism flips to reflective knowledge as specified in Sosa's virtue epistemology. AS3bv will go through the relevant levels of reflective processing within its own systems and with appropriate human input and control under IHL.

A demonstration of AS3bv is the way humans have designed the Tomahawk subsonic cruise missile to self-correct by comparing the terrain beneath it to satellite-generated maps stored on-board. If its trajectory is altered, motors will move the wings to counter unforeseen difficulties on its journey. The tactical Tomahawk can be reprogrammed mid-flight remotely to a different target using GPS coordinates stored locally. Tomahawk engineers' and operators' competencies play a role in the success or failure of the missile to hit its target. If part of the guidance system fails, human decisions will affect how well the missile flies. Part of the reason why credences need to play a greater role in epistemology is that instances where knowledge does not obtain—yet competent processes are deployed—should not prevent action toward a goal.

It is possible that a future LAWS may achieve reflective knowledge via a hierarchy of Bayesian processes, known as Hierarchically Nested Probabilistic Models (HNPM). HNPM are structured, describing the relations between models and patterns of evidence in rigorous ways emulating higher-order "Type 2," reflective capabilities (Devitt 2013). HNPM achieve higher-order information processing using iterations of the same justificatory processes that underlie basic probabilistic processes. HNPM show that higher-order theories (e.g., about abstract ideas) can become inductive constraints on the interpretation of lower-level theories or overhypotheses (Goodman 1983; Kemp et al. 2007). HNPM can account for multiple levels of knowledge, including (1) abstract generalizations relating to higher level principles, (2) specific theories about a set of instances, and (3) particular experiences. If the human mind is, to a great degree, Bayesian, then building LAWS that operate similarly may build trust, explainability, understandability, and better human-machine systems. AS3bv systems will be more virtuous because they will move with assurance in their actions, declare their uncertainties, reflect on their beliefs, and be constrained within operations according to their obligations under IHL and Article 36 guidelines.

Then the question is, at what threshold of virtue and competence would any group or authority actually release AS3bv into combat operations or into a war scenario? As wars are increasingly operating in Grey Zones, they are becoming a virtual and physical conflagration between private individuals, economic agents, militarized groups, and government agencies. The future of war will need agents operating in complicated social environments that require a defensible epistemology for how they make decisions. Combining Bayesian epistemology with virtue epistemology enables LAWS to operate rationally and cautiously in uncertain environments with partial and changing information. This level of defensible adaptability is important in emerging battlefront situations where the enemy is not clear, and individuals who need to be targeted must be carefully thought through.

In sum, knowledge is the ideal epistemic state, but epistemic states where knowledge falls short may leave behind another justified epistemic state—rational belief—plus a feedback opportunity to increase knowledge (or the probability of knowledge) for future situations. Reliable processes create knowledge, but also improve the odds of future knowledge in different conditions. Instances of knowledge are valuable in that they inform the agent and those around them of the scale of their competencies. Whether any LAWS passes an Article 36 review depends on human confidence that the weapon is reliable and competent in normal or expected use. This outcome depends on whether an autonomous weapon can be used in compliance with laws of armed conflict and act in a manner that is predictable, reliable, and explainable. A Bayesian virtue epistemology values reliability and competence and skills; knowledge and credences. Epistemology becomes not just the study of justified true belief then, but also the study of the processes of belief revision in response to confirming or disconfirming evidence. Justification arises from the apt performance of reliable processes and their coherence and coordination with other beliefs (Devitt 2013).

## Conclusion

This chapter has discussed higher-order design principles to guide the design, evaluation, deployment, and iteration of LAWS based on epistemic models to ensure that the lawfulness of LAWS is determined before they are developed, acquired, or otherwise incorporated into a States arsenal (International Committee of the Red Cross 2006). The design of lethal autonomous weapons ought to incorporate our highest standards for reflective knowledge. A targeting decision ought to be informed by the most accurate and fast information, justified over hierarchical levels of reliability enabling the best of human reasoning, compassion, and hypothetical considerations. Humans with meaningful control over LAWS ought to have knowledge that is safe, not lucky; contextually valid; and available for scrutiny. Our means of communicating the decision process, actions, and outcomes ought to be informed by normative models such as Bayesian and virtue epistemologies to ensure rational, knowledgeable, and ethical decisions.

Note